# Tunable phonon-induced transparency in bilayer graphene nanoribbons

Hugen Yan[1]*[‡], Tony Low[1]*[‡], Francisco Guinea[2], Fengnian Xia[1]*, and Phaedon Avouris[1]*

[1]IBM Thomas J. Watson Research Center, Yorktown Heights, NY 10598
[2]Instituto de Ciencia de Materiales de Madrid. CSIC. Sor Juana Inés de la Cruz 3. 28049 Madrid, Spain

**In the phenomenon of electromagnetically induced transparency[1] (EIT) of a three-level atomic system, the linear susceptibility at the dipole-allowed transition is canceled through destructive interference of the direct transition and an indirect transition pathway involving a meta-stable level, enabled by optical pumping. EIT not only leads to light transmission at otherwise opaque atomic transition frequencies, but also results in the slowing of light group velocity and enhanced optical nonlinearity[2]. In this letter, we report an analogous behavior, denoted as phonon-induced transparency (PIT), in AB-stacked bilayer graphene nanoribbons. Here, light absorption due to the plasmon excitation is suppressed in a narrow window due to the coupling with the infrared active Γ-point optical phonon[3,4], whose function here is similar to that of the meta-stable level in EIT of atomic systems. We further show that PIT in bilayer graphene is actively tunable by electrostatic gating, and estimate a maximum slow light factor of around 500 at the phonon frequency of 1580 cm$^{-1}$, based on the measured spectra. Our demonstration opens an avenue for the exploration of few-photon non-linear optics[5] and slow light[2] in this novel two-dimensional material, without external optical pumping and at**

**room temperature.**

†These authors contributed equally to the work.

*Emails: hyan@us.ibm.com (H.Y.), talow@us.ibm.com (T.L.), fxia@ us.ibm.com(F. X.) and avouris@us.ibm.com (P.A.)

Since the early demonstration of electromagnetically induced transparency (EIT) in atomic gases[1], analogous physical situations have been implemented in various solid state systems. This includes coupled optical resonators[6], metallic plasmonic structures[7-10] and opto-mechanical systems[11]. A plasmonic analogue of EIT utilizes the destructive interference effect between a radiative and a dark plasmon mode of different lifetimes[7]. A major motivation for the exploration of the EIT-like phenomenon in solid state systems is its potential in integrated photonic systems[6,12] for computing, optical communications, and bio-sensing[13], made possible by the enhanced light group index and nonlinearity within the spectral transparency window.

Graphene, with its unique relativistic-like linear energy dispersion, has emerged as a promising platform for plasmonics[14-18] due to its electrical tunability, strong light confinement, and relatively low plasmonic losses[18]. Very recently, spatially resolved propagating plasmons and tunable localized plasmons have been observed over a broad range of frequencies from the terahertz to the mid-infrared[14-17]. Besides graphene, several allotropes of carbon can also exhibit the above-mentioned attractive attributes for plasmonics. In this work, we focus on plasmons in bilayer graphene, showing that it is both an interesting and important plasmonic material in its own right.

The optical conductivity of AB-stacked bilayer graphene exhibits several interesting infrared features as revealed in recent measurements[3,4,19]. In particular, the coupling between the two graphene layers in bilayer graphene induces infrared activity on its Γ-point optical phonon, which exhibits a Fano-type[20] resonance in its infrared optical conductivity[3,4]. The Fano resonance is related to the presence of discrete states (i.e. phonon in this case) interacting coherently with a single particle continuum (electronic transitions). With the excitation of a different kind of quasi-particle, plasmon, we demonstrate an EIT-like behavior due to coherent interactions between the long-lived lattice vibration mode and the quasi-continuum plasmon mode of bilayer graphene nanoribbons. An extremely narrow transparency widow in the spectral response is observed, centered near the phonon frequency. To distinguish this from the widely studied plasmon-induced transparency[7-10,13], which typically involves two plasmon modes, we call the newly observed effect the "phonon-induced transparency" (PIT), a term that has been occasionally used in the literature under different circumstances[21]. This new effect modifies the infrared plasmonic response of bilayer graphene in drastic ways. At the spectral transparency, it is expected to be accompanied by sharp increase of the group index (or decrease of group velocity) and enhancement of the optical nonlinearity. A very wide degree of PIT tunability, both active and passive, is also demonstrated through electrostatic gating, chemical doping, and ribbon width control.

Our experimental results are found to be in good agreement with theoretical calculations, performed on a microscopic quantum mechanical footing.

The measurement scheme used in this study is shown in Fig. 1a. Large AB-stacked bilayer graphene flakes (> 100μm in size), were exfoliated from graphite through mechanical cleavage and identified by Raman spectroscopy[22] (Supplementary Information). The extinction spectra were obtained by comparing the transmission through the substrate with patterned bilayer graphene ribbons (or unpatterned bilayer graphene) and the bare substrate. Detailed sample preparation and characterization methods are presented in Methods. Fig. 1b shows an infrared extinction spectrum ($1-T/T_S$) of an unpatterned bilayer graphene. In this case, the transmission $T$ is independent of the light polarization. The first prominent feature in Fig. 1b is the broad extinction peak around 3500 cm$^{-1}$, which originates from the low energy electronic transitions[19], as shown in the lower inset. The second feature is the sharp phonon absorption peak at around 1580 cm$^{-1}$. The full width at half maximum (FWHM) of the peak is about 10 cm$^{-1}$. Previous studies have shown that the absorption magnitude and line-shape of the phonon depend strongly on the Fermi level and electrical field across the two graphene layers[3,4], which can be described by the charged phonon theory[23,24].

We patterned large area bilayer graphene into nanoribbons using electron beam lithography and reactive ion etching (see Methods). In this manner, plasmons can be excited using normal light incidence. Fig. 1c shows a scanning electron micrograph (SEM) of a bilayer graphene nanoribbon array. The extinction spectra of such an array with a ribbon width of 130 nm are shown in Fig. 2a, for light polarizations both perpendicular and parallel to ribbons. These two spectra are dramatically different due to the excitation of localized plasmon in the perpendicular polarization case[18], in which two prominent plasmon resonance peaks at around 1000 and 1400 cm$^{-1}$ are observed. These are coupled modes of plasmon and the surface polar phonon of the underlying $SiO_2$ substrate[18,25]. Throughout this work, we focus on the higher frequency plasmon mode (the one centered at ~1400 cm$^{-1}$ in Fig. 2a). Moreover, a strong graphene phonon peak with an extinction of > 5% exists at 1580 cm$^{-1}$. The extracted phonon peak as shown in the inset has a typical Fano line-shape[20], a result of interference between the sharp phonon resonance and the broad plasmon peak. The same phonon peak is also observed when the incident light polarization is parallel to the ribbons (grey curve) with much smaller extinction (< 2%). This large difference in the magnitude of the phonon extinction indicates that the coupling of external light to the phonon mode is enhanced significantly through plasmon excitation. This is analogous to that in plasmon-induced transparency, in which coupling of the external light excitation to the dark mode is enhanced through the radiative mode[7]. Moreover, this enhancement in phonon absorption also indicates that graphene can be utilized for surface enhanced infrared spectroscopy of molecular vibrations[26,27].

Decreasing ribbon width leads to an increase in the plasmon wave vector, resulting in an enhancement in plasmon resonance frequency, as previously demonstrated[18]. Figure 2b displays the extinction spectra of a ribbon array with width of 100 nm. When the plasmon frequency approaches the phonon frequency, the extinction spectrum exhibits a narrow transparency window at the phonon frequency, in sharp contrast to the case of light excitation with parallel polarization, in which an absorption peak shows up. This transparency window is due to the destructive interference of two optical transition pathways: excitations of a plasmon mode and a phonon mode. Compared to the typical plasmon-induced transparency where destructive interference of two plasmon modes are utilized, the PIT in bilayer graphene here has sharper transparency window due to the long phonon lifetime. This is desirable for applications such as slow light[2] and low-light level optical nonlinearity[5]. In conventional plasmon induced transparency, the dark plasmon mode, although longer lived than that of the radiative (bright) mode, still has very limited quality factor. Although utilization of superconductors as the plasmonic material can increase the dark mode's lifetime, since carrier scattering is suppressed in its superconducting state, it requires liquid helium temperature[10]. In this regard, bilayer graphene is a naturally superior material for this purpose, operating at room temperature.

We use a phenomenological theory involving two coupled classical oscillators[28] (as shown in the inset of Fig. 2b) to describe the PIT and the Fano resonance of the phonon feature. The details of the model are presented in the Supplementary Information. Solid line in Fig. 2b shows the fitted response, using a coupling strength of ~300 cm$^{-1}$ between the plasmon mode and the phonon mode. In this model, if the coupling strength is too large, the resulted extinction spectrum constitutes two well-separated modes and no narrow transparency window can be observed. For example, the hybridization of the graphene plasmon with $SiO_2$ surface polar phonon modes results in multiple, well-separated extinction peaks[18] (see spectra in Figs. 2a and 2c). On the other hand, if the plasmon-phonon coupling is too weak, the dip in the extinction spectrum will be small and such a small perturbation will not affect the group velocity and non-linear properties significantly (Supplementary Information). This is usually the case for the plasmon coupling to the molecular vibrations of attached molecules[26]. The coupling of plasmon to the infrared active phonon mode in bilayer graphene has the optimal strength such that a pronounced PIT effect can be observed. On the contrary, for thicker graphene sheets with three or four layers, we observe less pronounced PIT effect. This is probably due to the weaker plasmon-phonon coupling in those multilayer systems (see Supplementary Information).

In addition to the optimal plasmon-phonon coupling, the PIT in bilayer graphene is tunable. Figure 2c displays the extinction spectra for ribbon arrays with widths varying from 130 to 80 nm. The higher frequency plasmon peak can be tuned from below to above the phonon frequency. Fig. 2c clearly demonstrates the evolution process of the phonon line-shape, which varies from an enhanced Fano peak (130 nm ribbon) to a PIT-like absorption dip (100 nm ribbon) and finally a Fano peak again (80 nm ribbon).

Furthermore, the PIT can be tuned by doping as well. Figure 3a indicates the spectra for a ribbon array (100 nm) at different levels of chemical doping. The doping control procedure is detailed in the Methods. With increasing doping, the plasmon frequency up-shifts from below to above the phonon frequency. The extinction spectra evolve in a similar manner as that in Fig. 2c. Most importantly, the PIT in bilayer graphene can be actively controlled using electrostatic gating. Active control of conventional plasmon-induced transparency has been demonstrated recently through ultrafast laser excitations[8] and through temperature tuning of the superconducting elements[10]. We fabricated metal contacts on ribbons and gated the ribbons using a silicon back gate (see Methods), as illustrated in Fig. 1a. Fig. 3b presents the extinction spectra of a ribbon array (100 nm) with different back gate voltages. Again, PIT tunability is demonstrated. Here we want to emphasize that the group indices and non-linear properties associated with PIT are at the same time also tunable using gating, which may have significant impact on the future exploration of this bilayer plasmon-phonon system. A key parameter which describes the Fano line-shape of the phonon feature depends solely on the detuning of the plasmon frequency from that of the phonon. An analysis of the dependence is detailed in the Supplementary Information.

We have experimentally demonstrated a unique plasmonic system with bilayer graphene, where the interference between the plasmon and phonon modes leads to widely tunable Fano effect and PIT. Below, we present simulation results performed on a microscopic quantum mechanical level on this novel effect that allow new insights into the phenomenon. We consider a bilayer graphene arranged in the Bernal AB stacking order as depicted in the inset of Fig. 1b. Following McCann[29], we work in the $4\times 4$ basis of atomic $p_z$ orbitals (See Methods). The central quantity of interest is the dynamic dielectric function of the system $\varepsilon_T^{RPA}(q,\omega)$, which is calculated from the Random Phase Approximation (RPA). The coupling of the two optical in-plane phonons at $\Gamma$-point (i.e. the symmetric $E_g$ and the anti-symmetric $E_u$ modes) with the optically allowed electronic particle-hole transitions follows a formalism known as the charged-phonon theory[23,24], which accounts for the strong coupling between phonons and electronic transitions in an otherwise non-polar system like graphene. We defer further descriptions to the Supplementary Information.

The experimentally measured plasmon extinction spectrum is related to the RPA loss function $L(q,\omega) \equiv \text{Im}\left[1/\varepsilon_T^{RPA}\right]$ [30,31]. In this work, we use the simple mapping between plasmon momentum and the ribbon's width $q = \pi/(W-W_0)$, where $W_0$ denotes the width of the electrically dead zone[18]. Fig. 4a shows the calculated RPA loss spectra for bilayer graphene assuming the case for zero gap i.e. $\Delta = 0 eV$. In order to make comparison with experimentally measured extinction spectra, we employ in our simulations parameters accounting for known experimental conditions and knowledge acquired from prior work[18]: $q = \pi/(W-W_0) = 4.4\times 10^7 m^{-1}$ which correspond to ribbon array with $W = 100 nm$, $W_0 = 28 nm$, $T = 300K$, $\varepsilon_{env} = 1.5$, doping ranging from $\mu \approx -0.3 \rightarrow -0.4 eV$ at constant increment and phonon lifetime of $10 ps$. Finite electronic lifetimes is accounted for through the substitution $\hbar\omega \rightarrow \hbar\omega + i\eta$, where we assumed typical value of $\eta \approx 10 meV$. The qualitative agreement between our experimentally observed extinction spectra in Fig. 3a and the simulated result is satisfactory. In particular, the model describes well the evolution of the plasmon and infrared phonon resonances as they approach each other; going from separate resonances to the Fano-like asymmetric spectral line-shapes, and eventually an induced sharp transparency when their resonant frequencies coincide.

Fig. 4b shows an intensity plot of the loss function $L(q,\omega)$ in the vicinity of the phonon resonant frequency at 1580 cm$^{-1}$. With close to zero detuning, contrasting resonance linewidths and appropriate coupling strength between the two modes, destructive interference suppresses the absorption of the broader resonance, resulting in a very narrow transparency window. Fig. 4b also shows a giant transfer of spectral weight to the infrared phonon with decreased detuning, as reflected by the increase in both intensity and linewidth. The new elementary excitation leads to a "dressed" phonon with more pronounced infrared activity renormalized by many-body interactions. Comparison between the spectral weight of phonon mode with and without plasmon hybridization indicates a 100-fold enhancement in infrared activity, consistent also with experimental observation (see Fig. 2a). It is noteworthy that such plasmon-enhanced infrared absorption based on noble metals has produced an emerging field of spectroscopy techniques for surfaces and bio-molecules[27].

EIT is also known for its drastic modifications to the medium dispersion characteristics[2]. Recall that the group velocity describing propagation of wave packets can be expressed as $v_g = c/n_g$, where the group index is defined as $n_g \equiv n_r + \omega \cdot dn_r/d\omega$ with $n_r = \text{Re}[\sqrt{\varepsilon_T^{RPA}}]$ being the refractive index[2]. In the vicinity of the transparency, $\omega \cdot dn_r/d\omega$ can be significantly larger than $n_r$ in magnitude. Fig. 4c shows an intensity plot of the simulated group index, where $n_g$ can be as large as 500, or even negative in a narrow spectral window. The results indicate that bilayer graphene ribbons can potentially have dramatic effect on the propagation and interaction of infrared photons.

**Summary**

In summary, we have reported a novel phonon-induced transparency (PIT) phenomenon in the plasmonic response of bilayer graphene and demonstrated the wide tunability of this phenomenon, both passively and actively. Our microscopic theoretical model is in good agreement with the experimental observations, accounting for both the giant plasmonic enhancement of phonon infrared activity and the spectrally sharp transparency. In addition, PIT is usually accompanied by strong distortion in light dispersion, leading to a strong slow light effect. Our study therefore opens up a new avenue for EIT-like phenomenon in bilayer graphene metamaterials via its internal lattice vibration mode, and paves the way for various applications in few-photon non-linear optics, slow light devices, and biological sensing.

**Methods**

**Sample preparation, fabrication and measurement**

Graphene flakes on high resistivity $SiO_2$/Si substrate were mechanically exfoliated from graphite. The oxide thickness of the substrate is 90nm. Multilayer graphene flakes obtained in this way preserve good stacking order. Bilayer graphene, which is the focus of this paper, was identified by Raman and confirmed by infrared spectroscopy. Large area (>100μm in dimension) bilayer graphene flakes were chosen to make graphene nanoribbon arrays with area of 60μm ×60μm using electron beam lithography and oxygen plasma etching. The ribbon width was designed to be the same as or slightly larger than the gap between ribbons. For some of the ribbon arrays, Ti/Au metal contacts were also deposited to enable back gating.

The as-prepared ribbons are usually hole-doped (Fermi level µ<0). For the ribbons without metal contacts, the doping level can be increased further by exposing the samples to the nitric acid vapor for 10 minutes. The doping due to nitric acid can be removed partially or completely in ambient condition or by baking. As a consequence, we could achieve different doping levels. For the ribbons with metal contacts, we were able to actively change the Fermi level by a back gate.

The extinction measurements were done in a transmission geometry using a Nicolet 8700 FT-IR in conjunction with an IR-microscope. The IR beam size is ~25µm which is smaller than the ribbon array. To minimize the water absorption in the air, nitrogen gas was purged in the FT-IR chamber and the sample area. We measured the transmission $T_S$ through the bare area without graphene on the wafer as a reference and the transmission $T$ through the ribbon array with polarization either parallel or perpendicular to the ribbon axis. The extinction is defined as 1-$T/T_S$.

**Tight binding model for bilayer graphene**

In the band structure calculation, we consider a bilayer graphene arranged in the Bernal AB stacking order as depicted in the inset of Fig. 1b. Following McCann[29], we work in the $4\times 4$ basis of atomic $p_z$ orbitals ($a_1^\dagger, b_1^\dagger, a_2^\dagger, b_2^\dagger$) where $a_i^\dagger$ and $b_i^\dagger$ are creation operators for the i$^{th}$ layer on the sublattice A or B respectively. In this basis, the Hamiltonian near the **K** point can be written as:

$\hat{H}_k = v_f \pi_+ \hat{I} \otimes \hat{\sigma}_- + v_f \pi_- \hat{I} \otimes \hat{\sigma}_+ + (\Delta/2)\hat{\sigma}_z \otimes \hat{I} + (\gamma_1/2)[\hat{\sigma}_x \otimes \hat{\sigma}_x + \hat{\sigma}_y \otimes \hat{\sigma}_y]$, where $\hat{\sigma}_i$ and $\hat{I}$ are the $2 \times 2$ Pauli and identity matrices respectively, and we have defined $\hat{\sigma}_\pm = \frac{1}{2}(\hat{\sigma}_x \pm i\hat{\sigma}_y)$ and $\pi_\pm = \hbar(k_x \pm ik_y)$. Here, $v_f$ is the Fermi velocity, $\gamma_1$ the interlayer hopping and $\Delta$ is the electrostatic potential difference between the two layers. We derived the non-interacting ground state electronic bands and wavefunctions by diagonalizing the above Hamiltonian, see Supplementary Information for details.

**Acknowledgments**

The authors are grateful to H. Wang, W. Zhu, D. Farmer, M. Freitag, G. Tulevski, Y. Li, B. Ek and J. Bucchignano for experimental assistance in device fabrication and characterization. T. L. and F. G. acknowledge hospitality of KITP, supported in part by the NSF grant no. NSF PHY11-25915. T. L. also acknowledges partial support from NRI-INDEX and F.G. is also supported by the Spanish MICINN (FIS2008-00124, CONSOLIDER CSD2007-00010) and ERC grant 290846.

**Author contributions**

H.Y., F. X. and P. A. initiated the project and conceived the experiments. H.Y. and F.X. fabricated the devices. H.Y., F. X. performed the measurements and H. Y. analyzed the data. T.L. and F.G. provided modeling and theoretical foundations. P. A. provided advices throughout the project. H.Y. and T. L. co-wrote and all authors commented on the manuscript.

**Additional information**

**Figure captions**

**Figure 1. Schematics of the experiment**
**a**, Extinction spectrum measurement scheme for a gate-tunable bilayer graphene nanoribbon array. **b**, The extinction spectrum of an unpatterned bilayer graphene flake. The lower right inset depicts the low energy band structure of bilayer graphene with hole-doping ($\mu < 0$) and the dominant electronic transition responsible for the absorption peak is indicated. Upper left inset shows the lattice vibration responsible for the phonon absorption. **c**, A scanning electron micrograph of a typical graphene nanoribbon array used in the experiment. The scale bar is 200nm.

**Figure 2. Plasmon-phonon Fano system and Phonon-induced transparency in bilayer graphene nanoribbons**
**a**, Extinction spectra of a ribbon array with $W$=130nm for two incident light polarizations: parallel and perpendicular to the ribbons. The inset shows the extracted phonon spectrum with a Fano fit. **b**, Extinction spectra of a ribbon array with $W$=100nm. The spectrum for the perpendicular polarization is fitted by the coupled oscillator model, as shown by the solid curve. The inset depicts the coupled oscillator model scheme which is discussed in detail in the Supplementary Information. **c**, Ribbon width dependence of the spectra for the coupled plasmon-phonon Fano resonance system. Spectra are shifted vertically for clarity.

**Figure 3. Tunable phonon-induced transparency**
**a**, Spectrum evolution with increasing chemical doping for a ribbon array with $W$=100nm. **b**, Spectrum evolution with increasing back-gate voltage for a gated nanoribbon array device. Spectra are shifted vertically for clarity.

**Figure 4. Theoretical simulations of phonon-induced transparency and slow light**
**a**, Loss function, $L(q,\omega)$, of bilayer graphene simulated at particular $q=q_0$ corresponding to $W$=100nm. The different spectra (shifted vertically for clarity) are calculated at different Fermi level ranging from -0.3 to -0.4eV. See text for detailed simulation parameters. 2-dimensional intensity plots of $L(q,\omega)$ and group index $n_g(q,\omega)$ for the highest doping case, i.e. $\mu$ =-0.4eV, are shown in **b** and **c**. The value of $q_0$ is also indicated by the horizontal lines. Vertical dashed lines in **b** indicate the area of Fig. 4**c**.

**Figure 1 Schematics of the experiment**

a

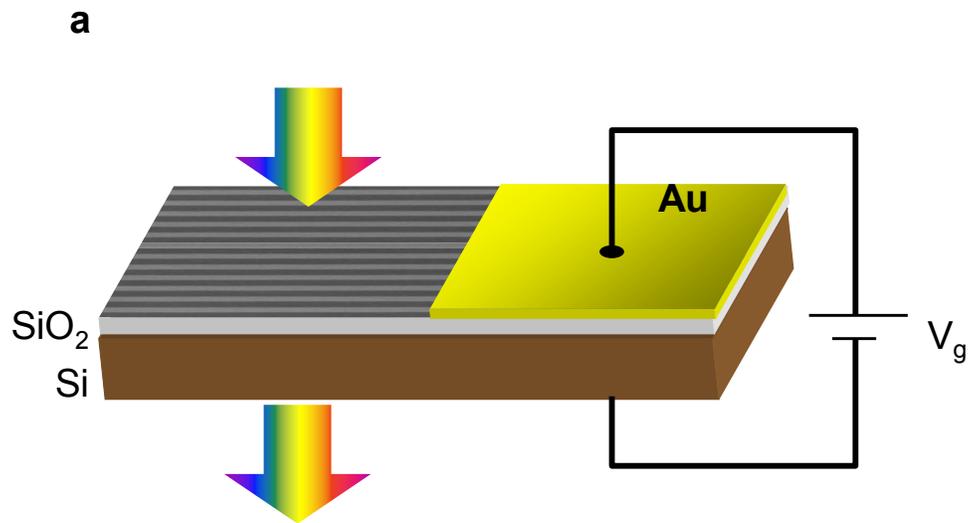

**b**

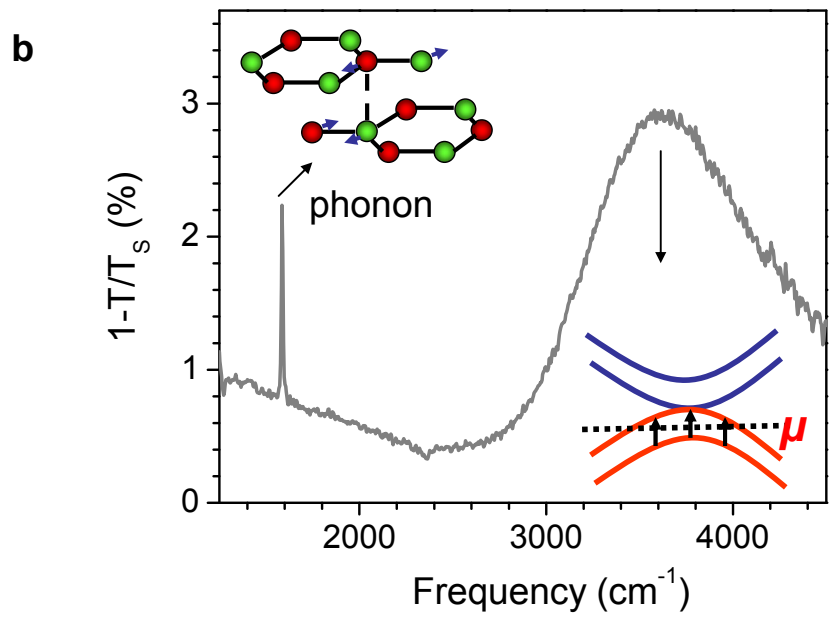

**c**

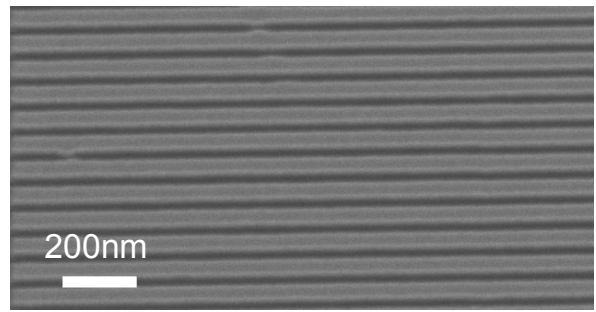

**Figure 2 Plasmon-phonon Fano system and Phonon-induced transparency in bilayer graphene nanoribbons**

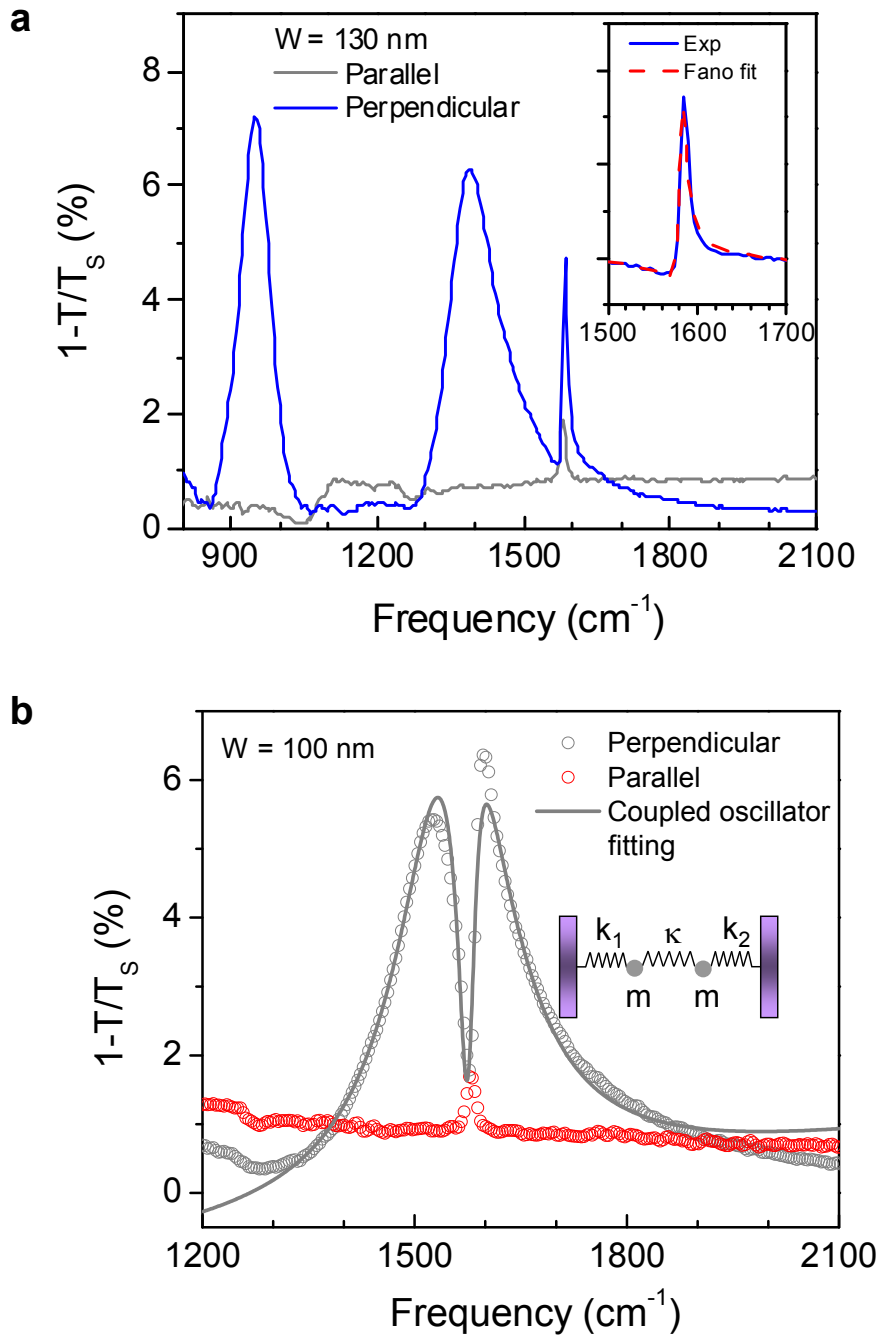

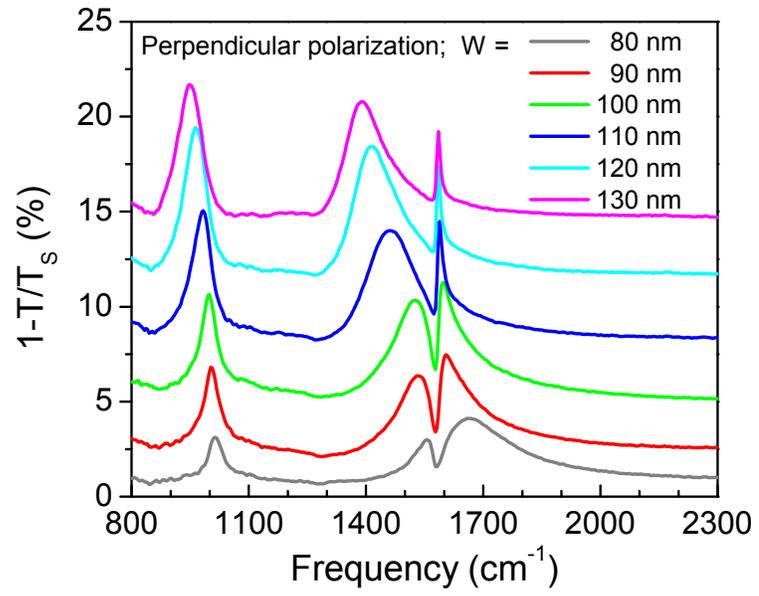

**Figure 3 Tunable phonon-induced transparency**

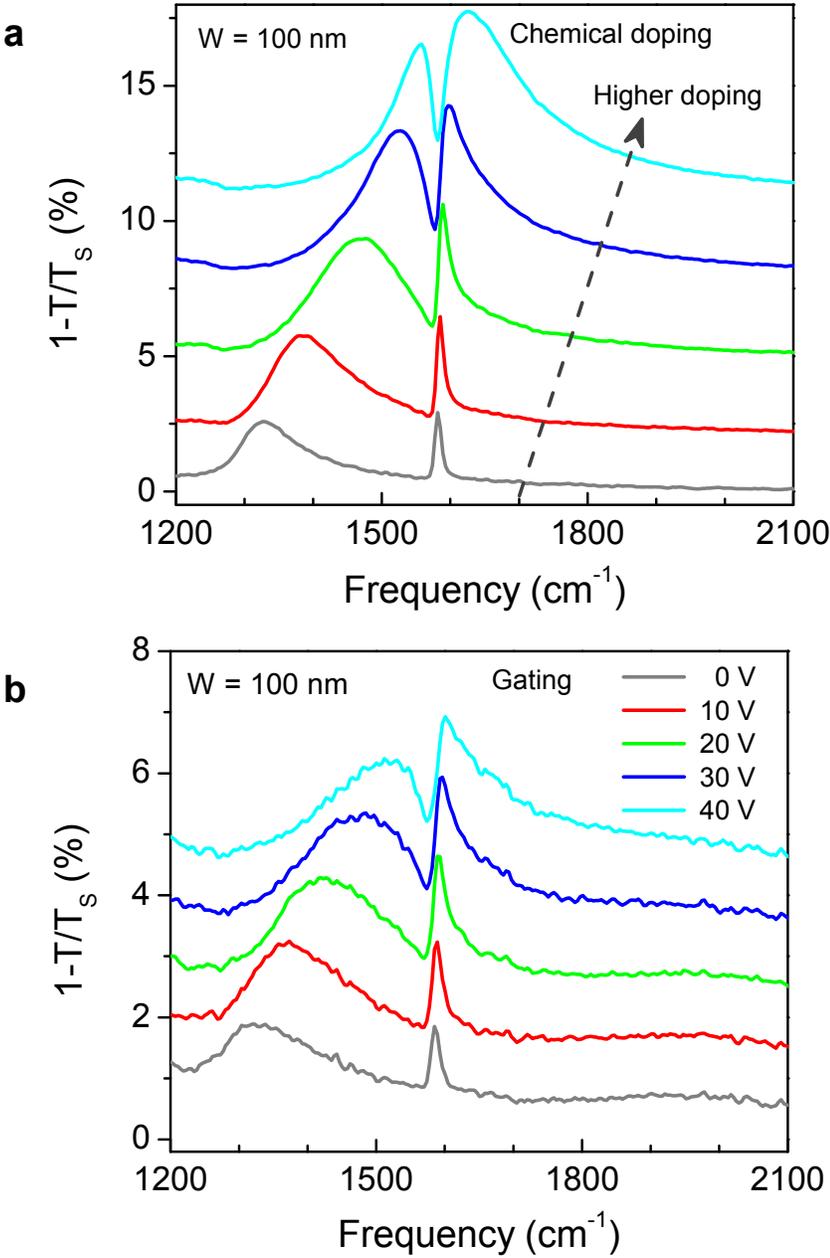

**Figure 4 Theoretical simulations of phonon-induced transparency and slow light**

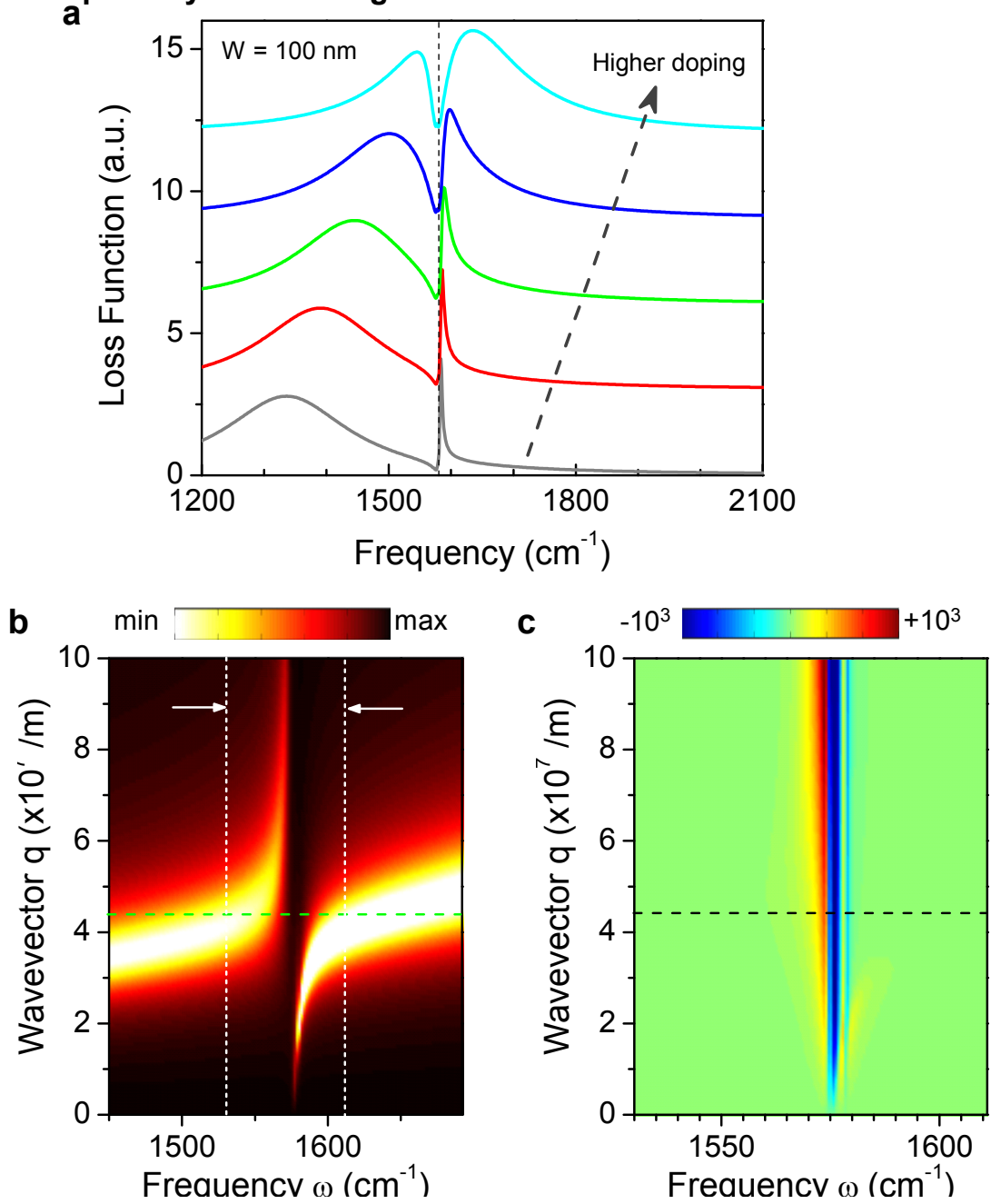